\documentclass[aps,twocolumn]{revtex4}

\usepackage{amsmath}
\usepackage{graphicx}

\begin{document}

%%%%%%%%%%%%%%%%%%%%%%%%%%%%%%%%%%%%%%%%%%%%%%%%%%%%%%%%%%%%%%%%%%%%%%%%%%%
%%%%%%%%%%%%%%%%%%%%%%%%%%%%%%%%%%%%%%%%%%%%%%%%%%%%%%%%%%%%%%%%%%%%%%%%%%%

\title{Pseudogap Kondo Physics from Charge 
 Fluctuations in a Quantum Dot}
\author{Eugene H. Kim~$^{1}$, Yong Baek Kim~$^{2}$, and C. Kallin~$^{1}$ }
\affiliation{$^1$ Department of Physics and Astronomy,
         McMaster University, Hamilton, Ontario, Canada L8S 4M1  \\
         $^2$ Department of Physics, University of Toronto, 
         Toronto, Ontario, Canada M5S 1A7}

\begin{abstract}
We consider charge fluctuations in a quantum dot coupled to
an interacting one-dimensional electron liquid.  We find the 
behavior of this system to be similar to the multichannel 
pseudogap Kondo model.  By tuning the coupling between the 
dot and the one-dimensional electron liquid, one can access 
the quantum critical point and the various fixed points 
which arise.  The differential capacitance is computed and 
is shown to contain detailed information about the system.
\end{abstract}
\maketitle

%%%%%%%%%%%%%%%%%%%%%%%%%%%%%%%%%%%%%%%%%%%%%%%%%%%%%%%%%%%%%%%%%%%%%%%%%%%
%%%%%%%%%%%%%%%%%%%%%%%%%%%%%%%%%%%%%%%%%%%%%%%%%%%%%%%%%%%%%%%%%%%%%%%%%%%

Nanotechnology has been the source of a renewed interest in 
the Kondo effect.\cite{leo}  The incredible progress in 
miniaturizing solid state devices has made it possible to 
fabricate small metallic islands ({\sl i.e. quantum dots})  
by confining electrons in a two-dimensional electron gas.  
Quantum dots provide a highly controllable environment to
study Kondo physics, and allow for many aspects of the 
Kondo effect to be probed.
In this work, we suggest that a quantum dot coupled to an 
interacting one-dimensional electron liquid ({\it i.e.} a 
Luttinger liquid) could provide a controlled environment
to observe pseudogap Kondo physics.

\begin{figure}
\scalebox{.52}{\includegraphics{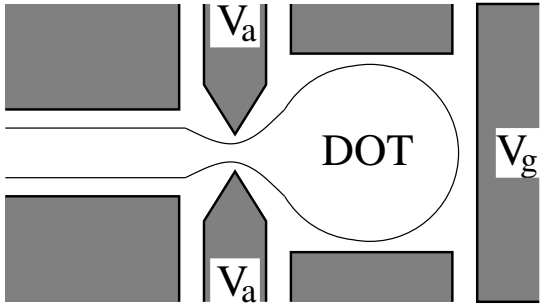} }
\caption{Setup: A quantum dot coupled to an interacting 
one-dimensional electron liquid.  The number of electrons 
on the dot is controlled by the gate voltage $V_g$.  A 
voltage applied to the auxiliary gates, $V_a$, controls 
the coupling between the dot and the one-dimensional
electron liquid.}
\label{fig:setup}
\end{figure}

The pseudogap Kondo model was first considered in 
Ref.~\onlinecite{withoff}.  In this model, a magnetic impurity 
is coupled to a sea of conduction electrons with a density 
of states vanishing at the Fermi energy with power-law 
behavior
\[
 H_{\rm int} = J~ {\bf \tau} \cdot \frac{ {\bf \sigma}_{s,s'}}{2}
 \psi^{\dagger}_{s}(0)\psi^{\phantom \dagger}_{s}(0) 
 \ \ \ {\rm with} \ \ \ \rho(E) = \rho_0 |E|^r \, . 
\]
One of the most interesting features of this model is that,
for antiferromagnetic coupling ($J>0$), there is an unstable
intermediate coupling fixed point occurring when $J = J_c$.
For $J < J_c$, the impurity spin is unscreened at low
energies; for $J > J_c$ the impurity spin is screened.
At $J = J_c$, the impurity spin exhibits quantum critical
fluctuations.
It is worth mentioning that this model has attracted attention
recently due to its potential relevance to various correlated
electron systems.  In particular, this model has been argued 
to describe impurities in high-$T_c$ cuprate superconductors.
\cite{zinc}  Moreover, the critical behavior occuring when 
$J=J_c$ may be relevant to the behavior seen in heavy fermion 
materials.\cite{si2,si1}

%%%%%%%%%%%%%%%%%%%%%%%%%%%%%%%%%%%%%%%%%%%%%%%%%%%%%%%%%%%%%%%%%%%%%%%%%%%
%%%%%%%%%%%%%%%%%%%%%%%%%%%%%%%%%%%%%%%%%%%%%%%%%%%%%%%%%%%%%%%%%%%%%%%%%%%

The setup we consider is shown in Fig.~\ref{fig:setup}.  
A large quantum dot is coupled to a reservoir, consisting 
of an interacting one-dimensional electron liquid.  The dot 
is capacitively coupled to a gate; the gate voltage $V_g$ 
controls the number of electrons on the dot.  The coupling 
between the dot and the reservoir is controlled by a voltage 
$V_a$ applied to the auxiliary gates.
To model the dot, we assume the level spacing of the dot is much 
smaller than any energy scale in the problem; we approximate the 
spectrum of the dot by a single particle continuum.
Moreover, we consider the case where the reservoir is coupled to 
the dot via a point contact.  
The Hamiltonian for the dot has the form 
$H_{\rm dot} = H_0({\rm dot}) + H_{\rm int}$.  
$H_0({\rm dot})$ describes the single particle energy levels of 
the dot; $H_{\rm int}$ describes the charging energy of the dot, 
as well as the coupling to the reservoir
\begin{equation}
 H_{\rm int} = \frac{E_c}{2}(\hat{N} - \overline{N})^2 
 + t \left(\psi^{\dagger}_{2,s}(0) \psi^{\phantom \dagger}_{1,s}(0) 
 + h.c. \right)  \, .
\label{Hdot}
\end{equation}
In Eq.~\ref{Hdot}, $\psi_{2,s}$ ($\psi_{1,s}$) destroys an 
electron in the dot (reservoir);  $\hat{N}$ is the number 
operator of the dot; $\overline{N}$ is the average number 
of electrons on the dot, which is proportional to $V_g$; 
$E_c$ is the charging energy of the dot; $t$ is the tunneling 
matrix element between the dot and the reservoir, which is 
controlled by $V_a$.
For generic values of $\overline{N}$, it costs a finite 
energy to put an extra electron on the dot; for temperatures 
sufficiently less that $E_c$, Coulomb blockade develops and 
the number of electrons on the dot becomes quantized.
However, for $\overline{N} = n + 1/2$ the energies of the 
$n$-electron and $(n+1)$-electron states are equal, and the 
charging energy vanishes.  Therefore, quantum fluctuations 
between the dot and the reservoir become important.

In the following, we assume $t$ is small and we focus on the 
regime $\overline{N} \approx n + 1/2$.  For energies sufficiently 
less than $E_C$, the physics will be dominated by the states 
$n$ and $(n+1)$.  Hence, we can project out all other states 
and restrict ourselves to this subspace.  By considering these 
states as the two states of a pseudospin $\tau^z = \pm 1/2$ 
and writing $\hat{N} = (n + 1/2) + \tau^z$, $H_{\rm int}$ takes 
the form\cite{matveev1}
\begin{equation}
 H_{\rm int} = t \left(\tau^+ \sigma^-_{i,j}
 \psi^{\dagger}_{i,s}(0) \psi^{\phantom \dagger}_{j,s}(0) + h.c. \right)
 - h \tau^z  \, ,
\label{effectivekondo}
\end{equation}
where $h = E_c [\overline{N} - (n+1/2)]$.
Eq.~\ref{effectivekondo} is a Kondo Hamiltonian with
anisotropic couplings.
Whereas the Kondo effect usually involves a magnetic impurity,
it arises in this system due to charge fluctuations.

Recently, it was argued that, besides the Kondo physics of
Eq.~\ref{effectivekondo}, other types of behavior are possible.
\cite{konik}  In particular, by performing a variational 
calculation, these authors identified that quantum fluctuations 
might give rise to tricritical Ising behavior.  However, it
remains to be seen whether these results will be confirmed 
numerically or experimentally.  In this work, we focus on 
regimes where the system is far from the potential tricritical 
point, so that the Kondo physics dominates.

Being interested in the low energy properties of the system, 
we expand the electron operator in the reservoir in terms of 
right and left movers
\[
\psi_{1,s}(x) = e^{ik_F x} \psi_{R,1,s}(x) 
    + e^{-ik_F x} \psi_{L,1,s}(x) \, ,
\]
where $k_F$ is the Fermi wavevector, and $\psi_{R,1,s}$ and 
$\psi_{L,1,s}$ are the (slowly varying) right and left moving 
fermion operators.
Moreover, upon expanding the electron operator in the dot in 
harmonics centered about the point contact, the reservoir couples 
to only a single harmonic.\cite{chamon}  Focussing on that single 
harmonic, we can write an effective one-dimensional model for the 
dot.\cite{chamon}  
In what follows, we will make extensive use of the boson 
representation.  To do so, the electron operator is written as 
$\psi_{R/L,i,s} \sim e^{\pm i \sqrt{4\pi} \phi_{R/L,i,s} }$
where the chiral fields, $\phi_{R,i,s}$ and $\phi_{L,i,s}$, 
are related to the usual Bose field $\phi_{i,s}$ and its dual 
field $\theta_{i,s}$ by $\phi_{i,s} = \phi_{R,i,s} + \phi_{L,i,s}$ 
and $\theta_{i,s} = \phi_{R,i,s} - \phi_{L,i,s}$.  It will 
also prove useful to form {\sl charge} and {\sl spin} fields
$\phi_{i,\rho/\sigma} = \left(\phi_{i,\uparrow} 
 \pm \phi_{i,\downarrow} \right)/\sqrt{2}$.
In terms of these variables, 
\begin{eqnarray}
 H({\rm lead}) & + & H_0({\rm dot})   \label{leadhamiltonian} \\
  & = & \frac{v_F}{2} \sum_{i=1}^2 \int_{-\infty}^0 \hspace{-.1in} dx~ 
  (\partial_x \theta_{i,\sigma})^2 + (\partial_x \phi_{i,\sigma} )^2 
  \nonumber \\  & + & 
  \frac{v_F}{2} \sum_{i=1}^2 \int_{-\infty}^0 \hspace{-.1in} dx~ 
   K_i~ (\partial_x \theta_{i,\rho})^2 
 + \frac{1}{K_i}~(\partial_x \phi_{i,\rho} )^2 
 \, . \nonumber
\end{eqnarray}
The Luttinger parameter in the reservoir, $K_1$, is determined by 
the interactions --- $K_1 < 1$ for repulsive interactions and 
$K_1 > 1$ for attractive interactions.  For the dot, $K_2 = 1$.
In this work, we will focus on the case of repulsive interactions,
$K_1 < 1$.
To analyze the physics it will prove useful to unfold the system,
and work solely in terms of right moving fields.\cite{eggert}
Moreover, by forming linear combinations of the Bose fields in 
the dot and the reservoir, the system can be treated as two 
identical Luttinger liquids with an effective Luttinger 
parameter\cite{chamon}
\begin{equation}
 K = \frac{2 K_1}{K_1 + 1}  \, .  
\end{equation}

%%%%%%%%%%%%%%%%%%%%%%%%%%%%%%%%%%%%%%%%%%%%%%%%%%%%%%%%%%%%%%%%%%%%%%%%%%%
%%%%%%%%%%%%%%%%%%%%%%%%%%%%%%%%%%%%%%%%%%%%%%%%%%%%%%%%%%%%%%%%%%%%%%%%%%%

\begin{figure}
\scalebox{.56}{\includegraphics{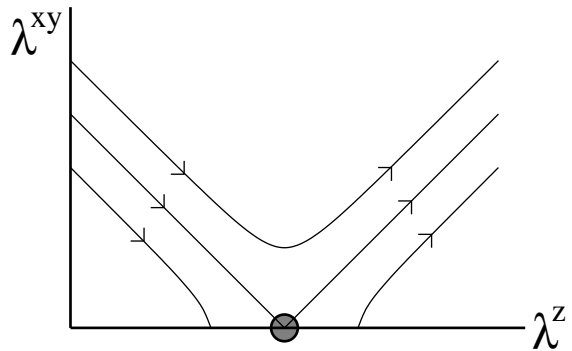} }
\caption{RG flows of Eq.~\ref{RGequations}. }
\label{fig:RGflows}
\end{figure}

The effects of Eq.~\ref{effectivekondo} can be deduced by a 
renormalization group (RG) analysis.  More generally, we will
consider 
\begin{eqnarray}
 H_{\rm int} & = & t \left(\tau^+ \sigma^-_{i,j}
 \psi^{\dagger}_{i,s}(0) \psi^{\phantom \dagger}_{j,s}(0) + h.c. \right)
 \nonumber \\
 & + & t'~ \tau^z \sigma^z_{i,j} \psi^{\dagger}_{i,s}(0) 
      \psi^{\phantom \dagger}_{j,s}(0)  - h \tau^z  \, .
\label{effectivekondomod}
\end{eqnarray}
Though the $t'$ term is not present in Eq.~\ref{effectivekondo}, 
it will be generated upon renormalization.  To lowest non-trivial
order, the RG equations for the parameters are
\begin{eqnarray}
 \frac{d\lambda^{xy}}{dl} =     
    \frac{1}{2}\left(1 - \frac{1}{K} \right) \lambda^{xy} 
    + \frac{1}{K} \lambda^{xy}\lambda^z  \ & , & \
 \frac{d \lambda^z}{dl} = \left(\lambda^{xy}\right)^2  \, , 
 \nonumber \\ & &  \hspace{-1.45in} 
 \frac{d \lambda^h}{dl} = \lambda^h - K (\lambda^{xy})^2 \lambda^h  
 \, . 
\label{RGequations}
\end{eqnarray}
where $\lambda^{xy} \sim t$, $\lambda^{z} \sim t'$, and 
$\lambda^{h} \sim h$.  The RG flows in the 
$\lambda^z - \lambda^{xy}$-plane are plotted in 
Fig.~\ref{fig:RGflows}.  Notice that there is a critical
point occurring when $\lambda^{xy}_c \equiv (1-K)/(2\sqrt{K})$.
For $\lambda^{xy} \leq \lambda^{xy}_c$ the coupling flows
to zero, while for $\lambda^{xy} > \lambda^{xy}_c$ the system 
flows to strong coupling.
Since this critical point arises in the same way as in the
pseudogap Kondo model, we will refer to it as the {\sl 
pseudogap Kondo critical point}.\cite{model}
For $\lambda^{xy} \leq \lambda^{xy}_c$, the system flows to 
the fixed point where the dot is decoupled from the reservoir.
(We will refer to this as the {\sl decoupled fixed point}.)
In terms of the effective Kondo model, the ``impurity'' is 
unscreened at low energies.
For $\lambda^{xy} > \lambda^{xy}_c$, $\lambda^{xy}$ initially decreases 
under the RG.  However, it will eventually start to increase and then 
flow off to strong coupling.  Integrating Eq.~\ref{RGequations}, we 
find that $\lambda^{xy} = {\cal O}(1)$ at a scale
\begin{equation}
 T_K = E_c \exp \left[\frac{-1}{|\delta|}\left[ 
 {\rm arccos}(|\delta|) - {\rm arctan}(x_0/|\delta|) \right] \right] \, ,
\label{TK}
\end{equation}
where $x_0 = (K-1)/(2K)$ and 
$|\delta| = \sqrt{(\lambda^{xy})^2/K - x_0^2}$. 
For energies below $T_K$, the dot and the reservoir are strongly 
coupled.  The strong coupling fixed point which arises is non-trivial 
--- it corresponds to the {\sl 2-channel Kondo fixed point} with 
a spin-1/2 impurity.\cite{matveev2}  This occurs because both 
spin-up and spin-down electrons in the reservoir try to occupy 
the single available charge state on the dot.

It should be noted that a related system was considered recently
in Ref.~\onlinecite{akira}.  In that work, the authors considered 
a resonant level coupled to a Luttinger liquid of spinless fermions.  
If the Luttinger parameter was smaller than some critical value, 
$K < K_c$, they too found a transition as one tuned the coupling 
between the dot and the Luttinger liquid.  The authors of 
Ref.~\onlinecite{akira} focussed on the zero temperature properties 
of their system.  In this work, we show that much rich physics 
can be observed at finite temperatures and frequencies.

%%%%%%%%%%%%%%%%%%%%%%%%%%%%%%%%%%%%%%%%%%%%%%%%%%%%%%%%%%%%%%%%%%%%%%%%%%%
%%%%%%%%%%%%%%%%%%%%%%%%%%%%%%%%%%%%%%%%%%%%%%%%%%%%%%%%%%%%%%%%%%%%%%%%%%%

The quantity of experimental interest is the differential 
capacitance.  In terms of the effective Kondo model, this
corresponds to the impurity susceptibility.
\cite{matveev1,matveev2}  Hence, we will need to calculate 
correlation functions of impurity operators.  To begin with, 
we will focus on the regime where $\lambda^{xy} < {\cal O}(1)$.  
In this regime, we can calculate the impurity susceptibility using
the RG.  In general, an N-point impurity correlation function
$G_N(\tau_1, \cdots \tau_N; \lambda_i, E_c) \equiv 
  \langle \tau^z(\tau_1) \cdots \tau^z(\tau_N) \rangle$
satisfies the RG equation
\begin{equation}
  \left[ \frac{\partial}{\partial l} 
 + \sum_i \beta_i \frac{\partial}{\partial \lambda_i}
 + N \gamma \right] G_N(\tau_1, \cdots \tau_N; \lambda_i, E_c) 
 = 0  \, ,
\label{greensRG}
\end{equation}
where $\beta_i = d \lambda_i / d l$, and $\gamma$ is the 
anomalous exponent.  The solution of Eq.~\ref{greensRG} is
\begin{eqnarray*}
 & & G_N(\tau_1, \cdots \tau_N; \lambda_i, E_c) =  
 \exp \left[-N \int_0^{l^*} dl~ \gamma(l) \right] 
 \nonumber \\  & & \hspace{.95in}  \times  
 G_N(\tau_1, \cdots \tau_N; \lambda_i(l^*), e^{-l^*} E_c) \, .
\end{eqnarray*}
Using Eq.~\ref{RGequations}, we obtain
\begin{eqnarray}
 & & G_N(\tau_1, \cdots \tau_N; \lambda_i, E_c) =  
 e^{-N K(1-K)/2} e^{-N K^2 f(l^*) }
 \nonumber \\  & & \hspace{.75in}  \times  
 G_N(\tau_1, \cdots \tau_N; \lambda_i(l^*), e^{-l^*} E_c) \, ,
\label{greensRGsolution}
\end{eqnarray}
where 
\begin{subequations}\label{RGsolutions}
\begin{eqnarray}
 f(l^*) & = & |\delta| \left[ \frac{ y_0^2
   + (|\delta| - x_0)^2 e^{2|\delta| l^* } }
  {y_0^2 - (|\delta| - x_0)^2 e^{2|\delta| l^* } }  \right] 
  \  (\lambda^{xy} < \lambda^{xy}_c)  
\label{RGsolutiona}  \\
 f(l^*) & = & \frac{x_0}{1-x_0~ l^* }   
  \ \ \ \ \ (\lambda^{xy} = \lambda^{xy}_c)   
\label{RGsolutionb}  \\
 f(l^*) & = & |\delta| \tan\left[ |\delta|~ l^* 
 + {\rm arctan}\left( \frac{x_0}{|\delta|} \right) \right] 
 \ (\lambda^{xy} > \lambda^{xy}_c) \, .  \nonumber \\
\label{RGsolutionc}
\end{eqnarray}
\end{subequations}
In Eq.~\ref{RGsolutions}, $x_0 = (K-1)/(2K)$;
$y_0 = \lambda^{xy}/\sqrt{K}$ and 
$|\delta| = \sqrt{x_0^2 - y_0^2}$ in Eq.~\ref{RGsolutiona};  
$|\delta| = \sqrt{(\lambda^{xy})^2/K - x_0^2}$ in 
Eq.~\ref{RGsolutionc}.
Choosing $e^{l^*} \sim E_c/E$, the correlation function on the 
right-hand-side of Eq.~\ref{greensRGsolution} can be evaluated 
perturbatively.

%%%%%%%%%%%%%%%%%%%%%%%%%%%%%%%%%%%%%%%%%%%%%%%%%%%%%%%%%%%%%%%%%%%%%%%%%%%
%%%%%%%%%%%%%%%%%%%%%%%%%%%%%%%%%%%%%%%%%%%%%%%%%%%%%%%%%%%%%%%%%%%%%%%%%%%

\begin{figure}
\scalebox{.31}{\includegraphics{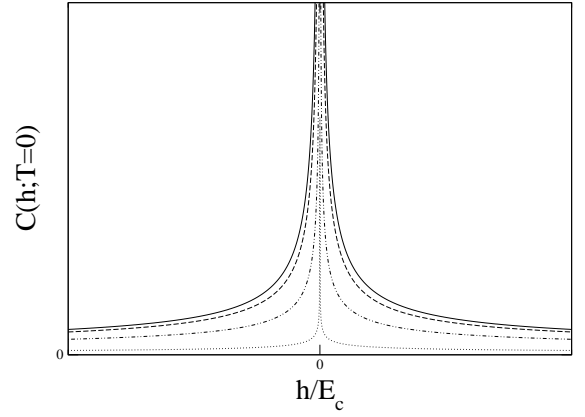} }
\caption{$C(h;T=0)$ vs. $h$ for 
$\lambda^{xy} \leq \lambda^{xy}_c$ (with $K=0.6$).  
Solid line: $|\delta|=0$; dashed line: $|\delta|=0.1$; 
dash-dotted line: $|\delta|= 0.2$; dotted line: 
$|\delta|=0.3$.}
\label{fig:capgateplot}
\end{figure}

We start by considering the temperature dependence of the 
differential capacitance on resonance, $C(h \rightarrow 0;T)$.  
Using Eq.~\ref{greensRGsolution} with $N=2$, we obtain
\begin{equation}
 C(h \rightarrow 0;T) = \frac{ e^{K(K-1)} }{4T}~ e^{-2 K^2 f(T)} \, ,
\label{RGimprovedT}
\end{equation} 
where $f(T)$ is given by Eq.~\ref{RGsolutions} with 
$l^* = \ln (E_C/c_1 T)$.  ($c_1$ is an ${\cal O}(1)$ constant.)
From Eq.~\ref{RGimprovedT}, we see that 
$C(h\rightarrow 0;T) \sim T^{-1}$ near the decoupled fixed 
point ($T \rightarrow 0$ for $\lambda^{xy} \leq \lambda^{xy}_c$).  
Moreover, near the pseudogap Kondo critical point 
($\lambda^{xy} = \lambda^{xy}_c$ for $T \gg E_c \exp[2K/(K-1)]$)
$C(h\rightarrow 0;T) \sim T^{\Delta_T-1}$ where 
$\Delta_T = (1-K)^2/2$.
It is also interesting to consider the differential capacitance 
at $T=0$ as a function of gate voltage, 
$C(h;T=0) = E_c~ d \langle \tau^z \rangle / dh$.
Using Eq.~\ref{greensRGsolution} with $N=1$, we obtain
\begin{equation}
 \langle \tau^z \rangle = \pm \frac{e^{K(K-1)/2}}{2} 
  e^{-K^2 f(|h|)} \, ,
\label{RGimprovedh}
\end{equation} 
where the plus (minus) sign is for $h > 0$ ($h < 0$), and $f(|h|)$ 
is given by Eq.~\ref{RGsolutions} with $l^* = \ln(E_C/c_2 |h|)$.
($c_2$ is another ${\cal O}(1)$ constant.)  
The differential capacitance vs. gate voltage is plotted in 
Fig.~\ref{fig:capgateplot}.  Near the decoupled fixed point,
we find $C(h;T=0) \sim |h|^{2|\delta| - 1}$ as $|h| \rightarrow 0$
($|\delta|$ is as in Eq.~\ref{RGsolutiona}). Also, near the pseudogap 
Kondo critical point $C(h;T=0) \sim |h|^{\Delta_h-1}$ where 
$\Delta_h = (1-K)^2/4$ ($|h| \gg E_c \exp[2K/(K-1)]$).

Notice that near the decoupled fixed point,
$C(h\rightarrow 0;T) \sim T^{-1}$ as $T\rightarrow 0$.  This 
Curie-Weiss-like form arises because the ``impurity'' behaves 
basically like a free spin.  However, the local ``moment'' is 
reduced from its non-interacting value.\cite{vojta,konik,akira}  
From Eqs.~\ref{RGimprovedh}, we see that the amount by which 
the local ``moment'' is reduced depends on $K$, as well as the 
value of $\lambda^{xy}$.
Also, notice that $\Delta_T = 2 \Delta_h$ near the pseudogap 
Kondo critical point.  In Ref.~\onlinecite{si1}, it was shown 
that the exponents near the critical point of the pseudogap 
Kondo model satisfy certain hyperscaling relations.  To the 
order of accuracy that we have worked, our results are consistent 
with these hyperscaling relations.

%%%%%%%%%%%%%%%%%%%%%%%%%%%%%%%%%%%%%%%%%%%%%%%%%%%%%%%%%%%%%%%%%%%%%%%%%%%
%%%%%%%%%%%%%%%%%%%%%%%%%%%%%%%%%%%%%%%%%%%%%%%%%%%%%%%%%%%%%%%%%%%%%%%%%%%

Now, we consider the physics in the regime 
$\lambda^{xy} > \lambda^{xy}_c$ for energies below $T_K$
(with $T_K$ given by Eq.~\ref{TK}).  In this regime, the 
system is close to the 2-channel Kondo fixed point.  To  
proceed, we follow Ref.~\onlinecite{schiller} and form 
combinations of the fields in the dot and the reservoir: 
$\phi_{R,c}$, $\phi_{R,sp}$, $\phi_{R,f}$, and $\phi_{R,sf}$.  
Then, we perform the unitary transformation,
$U = \exp \left( i \sqrt{4\pi / K}~ \tau^z \phi_{R,f}(0) \right)$,
which ties charge to the ``impurity''.
Finally, we introduce new fermion fields, $d \sim \tau^-$,
$X \sim e^{i\sqrt{4\pi}\phi_{R,sf}}$, and 
$f \sim e^{i\sqrt{4\pi}\phi_{R,f}}$.
Upon performing these transformations, $H_{\rm int}$ becomes
\begin{eqnarray}
 H_{\rm int} & = & v_F \tilde{\lambda}^{xy} 
   \left( d^{\dagger} + d~ \right) \left( X^{\dagger}(0) - X(0) \right)     
 \nonumber \\   
 & + & v_F \sqrt{4\pi /K} \left(\tilde{\lambda}^{z} - 1 \right) 
   \left( d^{\dagger} d^{\phantom \dagger} - 1/2 \right) 
   f^{\dagger}(0) f(0)  \nonumber \\ 
 & - & v_F \tilde{\lambda}^h \left( d^{\dagger} d^{\phantom \dagger} 
   - 1/2 \right) \, , 
\label{toulouse}  
\end{eqnarray}
where $\tilde{\lambda}^{xy}$, $\tilde{\lambda}^{z}$, and 
$\tilde{\lambda}^h$ are the renormalized values of the couplings.  

Using Eq.~\ref{toulouse}, we can calculate the differential 
capacitance near the 2-channel Kondo fixed point.  Starting 
with the differential capacitance on resonance, we find 
(ignoring the irrelevant $(\tilde{\lambda}^{z} -1)$ term)
\begin{eqnarray}
 & & C(h \rightarrow 0; \omega,T) =  \label{2channelchi} \\
 & & \hspace{.2in} \frac{1}{T_K} \int \frac{dx}{2\pi}
  {\rm tanh}\left(\frac{x T_K}{2T}\right)
  \frac{1}{x^2 + 1} \frac{1}{x - (\omega/T_K) - i0^+}  \, , 
 \nonumber
\end{eqnarray}
For $\omega=0$ and $T \ll T_K$, this reduces to the
well-known result for the impurity susceptibility of the
2-channel Kondo model
$ C(h \rightarrow 0;T) = 1/(\pi T_K) \ln (T_K/T)$.
We can also calculate $C(h;T=0)$.  Using Eq.~\ref{toulouse}
(ignoring the irrelevant $(\tilde{\lambda}^z -1)$ term)
\begin{equation}
 \langle \tau^z \rangle = \frac{h}{T_K} 
 \int \frac{dx}{2\pi} {\rm tanh}\left(\frac{xT_K}{2T}\right)
 \frac{x}{(x^2 - (h/T_K)^2)^2 + x^2 }  \, .
\end{equation}
For $T = 0$ and $|h| \ll T_K$,
$C(h;T=0) = 4/(\pi T_K) \ln (T_K/|h|)$.
Notice that $C(h\rightarrow 0; T)$ ($C(h;T=0)$) diverges as 
$T \rightarrow 0$ ($|h| \rightarrow 0$).  However, the divergence
in this case is weaker than what occurs near the decoupled fixed 
point.  This is because, near the 2-channel Kondo fixed point, 
charge is tied to the ``impurity''.  As a result, the ground 
states $\tau^z = 1/2$ and $\tau^z = -1/2$ are orthogonal, in 
that they are not connected by $\tau^+$ or $\tau^-$.\cite{giamarchi}  
This removes the power-law divergence which occurs near the 
decoupled fixed point, and replaces it with the weaker logarithmic 
divergence.

%%%%%%%%%%%%%%%%%%%%%%%%%%%%%%%%%%%%%%%%%%%%%%%%%%%%%%%%%%%%%%%%%%%%%%%%%%%
%%%%%%%%%%%%%%%%%%%%%%%%%%%%%%%%%%%%%%%%%%%%%%%%%%%%%%%%%%%%%%%%%%%%%%%%%%%

\begin{figure}
\scalebox{.4}{\includegraphics{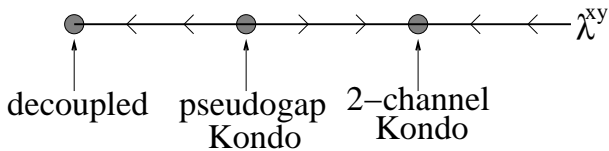} }
\caption{Fixed points and RG flows for the pseudogap Kondo
model.}
\label{fig:globalflows}
\end{figure}

In the above discussion, we saw three fixed points arise 
(shown schematically in Fig.~\ref{fig:globalflows}): (1) 
the decoupled fixed point, (2) the pseudogap Kondo critical
point, and (3) the 2-channel Kondo fixed point.  The pseudogap
Kondo critical point and the 2-channel Kondo fixed points 
are particularly interesting because they are non-trivial 
scale invariant fixed points.  As a consequence, one should 
be able to observe $\omega /T$-scaling near these fixed points 
by applying an AC component to the gate voltage.  
More specifically, we expect the dynamical capacitance on 
resonance to have the form
\begin{equation} 
 C(h \rightarrow 0;\omega,T) = T^{\nu - 1} X(\omega/T)  \, .
\label{omegaT}
\end{equation}
Near the pseudogap Kondo critical point, we can calculate 
the scaling function $X(\omega/T)$ for $(1-K) \ll 1$.  In 
the leading logarithm approximation, we find $\nu = \Delta_T$ 
($\Delta_T = (1-K)^2/2$) and 
\[
 X(\omega/T) = \frac{1}{4\pi}\left(\frac{2\pi}{E_c}\right)^{\nu}
   \sin\left(\frac{\pi \nu}{2}\right)
   B\left(\frac{\nu}{2} - i \frac{\omega}{2\pi T} , 
   1 -\nu \right)
\]
where $B$ is the beta function.
Near the 2-channel Kondo fixed point, we use Eq.~\ref{2channelchi}
to obtain $\nu = 1$ and 
\[
 X(\omega /T) = \frac{1}{\pi T_K} \ln \left( 
   \frac{T_K}{{\rm max}(\omega,T)} \right)
 + \frac{i}{2 T_K}~ {\rm tanh}\left(\frac{\omega}{2T}\right) \, .
\]
Note that $X(\omega /T)$ is, in general, complex.  Therefore, 
the differential capacitance will have components both in-phase 
and out-of-phase with the gate voltage.

%%%%%%%%%%%%%%%%%%%%%%%%%%%%%%%%%%%%%%%%%%%%%%%%%%%%%%%%%%%%%%%%%%%%%%%%%%%
%%%%%%%%%%%%%%%%%%%%%%%%%%%%%%%%%%%%%%%%%%%%%%%%%%%%%%%%%%%%%%%%%%%%%%%%%%%

To summarize, a (large) quantum dot coupled to an interacting
one-dimensional electron liquid could provide a controlled
environment to observe pseudogap Kondo physics.  By tuning the 
coupling between the dot and the one-dimensional electron 
liquid, one can access the various fixed points which arise:
the decoupled fixed point, the pseudogap Kondo critical point,
and the 2-channel Kondo fixed point.  Moreover, this system 
provides the remarkable opportunity to directly probe impurity 
properties via differential capacitance measurements.  
As the differential capacitance of a large quantum dot has 
recently been measured,\cite{ashoori} we are hopeful that 
the physics described in this work can be observed in the 
near future.

%%%%%%%%%%%%%%%%%%%%%%%%%%%%%%%%%%%%%%%%%%%%%%%%%%%%%%%%%%%%%%%%%%%%%%%%%%%
%%%%%%%%%%%%%%%%%%%%%%%%%%%%%%%%%%%%%%%%%%%%%%%%%%%%%%%%%%%%%%%%%%%%%%%%%%%

We would like to thank G. Sierra and A. Furusaki for very
helpful discussions.  
This work was supported by the NSERC of Canada (EHK, YBK, CK),
Materials and Manufacturing of Ontario (EHK, YBK, CK), Canada 
Research Chair (YBK), and Alfred P. Sloan Foundation (YBK).

%%%%%%%%%%%%%%%%%%%%%%%%%%%%%%%%%%%%%%%%%%%%%%%%%%%%%%%%%%%%%%%%%%%%%%%%%%%
%%%%%%%%%%%%%%%%%%%%%%%%%%%%%%%%%%%%%%%%%%%%%%%%%%%%%%%%%%%%%%%%%%%%%%%%%%%

%%%%%%%%%%%%%%%%%%%%%%%%%%%%%%%%%%%%%%%%%%%%%%%%%%%%%%%%%%%%%%%%%%%%%%%%%%%
%%%%%%%%%%%%%%%%%%%%%%%%%%%%%%%%%%%%%%%%%%%%%%%%%%%%%%%%%%%%%%%%%%%%%%%%%%%


\begin{thebibliography}{100}
\bibitem{model}{Our model is not identical to the pseudogap 
Kondo model.  In the pseudogap Kondo model, the $x$, $y$, and 
$z$ components of the Kondo interaction scale in the same way.  
In our model, the $x$ and $y$ components scale differently from 
the $z$ component.  (See Eq.~\ref{RGequations}.)}
\end{thebibliography}
\end{document}